\documentstyle[epsfig]{lamuphys}
\makeatletter
\let\chapter\hid@chapter
\makeatother
\newcommand{\be}{\begin{equation}}
\newcommand{\ee}{\end{equation}}
\newcommand{\ba}{\begin{eqnarray}}
\newcommand{\ea}{\end{eqnarray}}

\newcommand{\IcA}{{\Im m\cal A}}
\newcommand{\RcA}{{\Re e\cal A}}
\newcommand{\Frac}[2]{\frac{\displaystyle #1}{\displaystyle #2}}
\newcommand{\kpiggp}{$K^+ \rightarrow \pi^+ \gamma \gamma $ }
\newcommand{\kpiggtot}{$K \rightarrow \pi \gamma \gamma $ }

\newcommand{\kpiggn}{$K_L \rightarrow \pi^0 \gamma \gamma $ }

\newcommand{\kgll}{$K_L \rightarrow  \gamma \ell^+ \ell^- $ }

\newcommand{\kggs}{$K_L \rightarrow  \gamma \gamma^* $ }
\newcommand{\klmm}{$K_L \rightarrow  \mu^+ \mu^- $ }
\newcommand{\opc}{${\cal O}(p^4)$ }
\newcommand{\ops}{${\cal O}(p^6)$}
\newcommand{\ov}{\overline}
\begin{document}
\pagenumbering{arabic}
\titlerunning{Report of the Working Group on Goldstone Boson Production
and Decay}
\title{\hfill{\large LU TP 97/31}\\[-0.2cm]
\mbox{ } \hfill{\large hep-ph/9710555}\\[-0.2cm]
\mbox{ } \hfill{\large October 1997}\\
Report of the Working Group on Goldstone Boson Production
and Decay}

\author{J.~Bijnens\inst{1}, G.~Colangelo\inst{2}, G.~D'Ambrosio\inst{3},
F.~Gabbiani\inst{4}, G.~Isidori\inst{2},\\
J.~Kambor\inst{5},
A.~Kupsc\inst{6}, M.~Moinester\inst{7}, B.~Nefkens\inst{8},
J.~Oller\inst{9},
E.~Pallante\inst{10}, J.R.~Pel\'aez\inst{11}, B.~Renk\inst{12}
and P.~Talavera\inst{13}}

\institute{Convener, Department of Theoretical Physics 2, University of Lund,\\
S\"olvegatan 14A, S22362 Lund, Sweden
\and
INFN, Laboratori  Nazionali di Frascati,\\
P.O. Box 13, I-00044 Frascati, Italy
\and
INFN-Sezione di Napoli and Dipartimento di Scienze Fisiche,\\
Universit\`a di Napoli, I-80125 Naples, Italy
\and
    Dept. of Physics, Duke University,\\
    Durham, NC 27708, USA
\and
Theoretische Physik, Universit\"at Z\"urich,\\
CH 8001 Z\"urich, Switzerland
\and
Department of Radiation Sciences, University of Uppsala,
Box 535,\\
S-75121 Uppsala, Sweden; representing WASA/PROMICE Collaboration
\and
School of Physics and Astronomy, Univ. of Tel Aviv,\\
Tel Aviv 69978, Israel
\and
    Dept. of Physics, University of California,\\
    Los Angeles, CA 90024, USA
\and
Dept. de F\'{\i}sica Te\'orica and I.F.I.C.
Universidad de Valencia - C.S.I.C.\\
46100 Burjassot (Valencia) - Spain
\and
Inst. f. Theoretische Physisk, Universit\"at Bern\\
Sidlerstr. 5, CH-3012 Bern, Switzerland
\and
Dept. de F\'{\i}sica Te\'orica\\
Universidad Complutense. 28040 Madrid. Spain
\and
Inst. f. Physik, Universit\"at Mainz,\\
D-55099 Mainz, Germany
\and
Departament de F\'isica i Enginyeria Nuclear,\\
Universitat
Polit\`ecnica de Catalunya, E-08034 Barcelona, Spain
}

\authorrunning{J. Bijnens et al.}
\maketitle

\begin{abstract}
This is the summary of the working group on Goldstone Boson Production and
Decay of the Chiral Dynamics Workshop in Mainz, September 1-5, 1997. For
the production aspects we discuss $\pi^0$ and $\eta$ production in
nucleon-nucleon collisions and the behaviour of hadrons in various sum rules.
For the decays we present a discussion on various $K$, $\eta$ and $\pi$
decay channels.
Other aspects discussed are a new
treatment of meson-meson scattering, the light-by-light contribution to
the muon anomalous magnetic moment and progress in various aspects
of the $p^6$ generating functional in the mesonic sector.
\end{abstract}
\section{Introduction}
Chiral symmetry plays a prominent role in low energy hadronic physics.
The workshop was basically fully devoted to theoretical and experimental
progress in this area. In our working group we discussed some
topics not covered by the other three working groups (\cite{WG1,WG2,WG3})
and in the various plenary talks. The subject of chiral symmetry and the
modern way to extract its consequences, Chiral Perturbation Theory (CHPT),
was introduced by J\"urg Gasser (\cite{Gasser}). In this
working group we did not discuss the extension to a small quark condensate
(\cite{Stern} and references therein). The plan of this contribution follows
closely the summary as presented by the convener J.~Bijnens. In particular
some of the contributions to this working group are discussed in
his plenary talk (\cite{Bijnens}) and will not be repeated here.

On the first day we discussed the experiments done by WASA/PROMICE at
Celsius in Uppsala (A.~Kupsz) on meson productions and their future
$\eta$ decay program, as well as new $\eta$ decay results and future
programs (B.~Nefkens) and the relevant CHPT aspects including the
anomaly (J.~Bijnens).
The subject of anomalies was also covered experimentally (M.~Moinester).
The present status of theory for the anomalous magnetic moment of the
muon and its relevance for beyond the standard model effects were
covered. The main remaining uncertainty in the prediction
remains the low-energy hadronic cross-section (E.~Pallante). The subject
of chiral sum rules in the $VV$ and $AA$ sector has recently had a boost
by the two-loop calculations in CHPT (J.~Kambor). Further progress
at this order in CHPT was evident in the entire workshop. The topics
covered here were $\pi\to e\nu\gamma$ (P.~Talavera) and
various subtleties in the general $p^6$ calculations as well as first results
on the infinity structure at $p^6$ (G.~Colangelo).

Beyond CHPT there is the discussion of the generalization of the
Lippman-Schwinger approach and the Inverse Amplitude Method applied to
meson-meson scattering (J.~Pel\'aez) and the attempts to calculate
relevant higher order terms using various models, double VMD or not
was a recurring
theme in both $\eta$ and $K$ decays and the muon anomalous magnetic moment.

Experimental results in $K$ decays and future prospects were reviewed (B.~Renk)
as well as the KLOE program at DA$\Phi$NE (G.~Colangelo). Theoretical
developments were discussed in the both the main decay modes
(G.~Isidori) as in various rare ones (G.~D'Ambrosio, F.~Gabbiani).

This report is organized as follows in Sect. \ref{productionI} we
discuss various $\eta$ and $\pi$ production experiments and future plans here.
This also includes the anomalous process $\gamma\pi\to\pi\pi$.
In Sect. \ref{decays} we discuss the parts mentioned above relevant
for $\eta$ and $K$ decays.
Here we also review the uncertainty on the decay $\pi^0\to\gamma\gamma$.
Sect. \ref{productionII} contains the discussion of
the sum rules for $g-2$ and $VV$ and $AA$ as well as the
meson-meson scattering by the IAM.
The last section, Sect. \ref{other}, covers the progress in the $p^6$
generating functional.

The topics discussed in this working group but covered in the plenary talk
by J.~Bijnens (\cite{Bijnens}) are: radiative $\pi^+$ decay and the
light-by-light contribution to the muon anomalous magnetic moment.
\section{Production I}
\label{productionI}
\subsection{Hadronic Production of $\pi$ and $\eta$}
In Uppsala, an experimental programme for near threshold $\eta$ and $\pi$ meson
production in proton-nucleon interactions is carried out.  The experiments are
done at the CELSIUS accelerator storage ring using an internal cluster gas-jet
target and the WASA/PROMICE detector capable of measuring forward going
charged particles and decays photons from the neutral mesons (\cite{CA+96a}).
The photon detection allows for measurements very close to threshold when the
outgoing protons or deuterons escape undetected in the beam pipe.  At higher
energies, a kinematically complete reconstruction of the events is possible.

An example is the near-threshold measurement of the reaction $pp\rightarrow pp
\pi^\circ$ that has been measured at seven different excess energies, from 0.5
MeV to 14 MeV in the centre-of-mass system (\cite{BO+95}). The experiment gives
added weight to an interpretation that the reaction near threshold is sensitive
to the short range components of the nucleon-nucleon interaction.  The
understanding of the cross section magnitude remains a challenge for CHPT
(\cite{Kolck}).

\begin{figure}
\begin{center}
\mbox{\epsfig{height=8cm,width=9cm,file=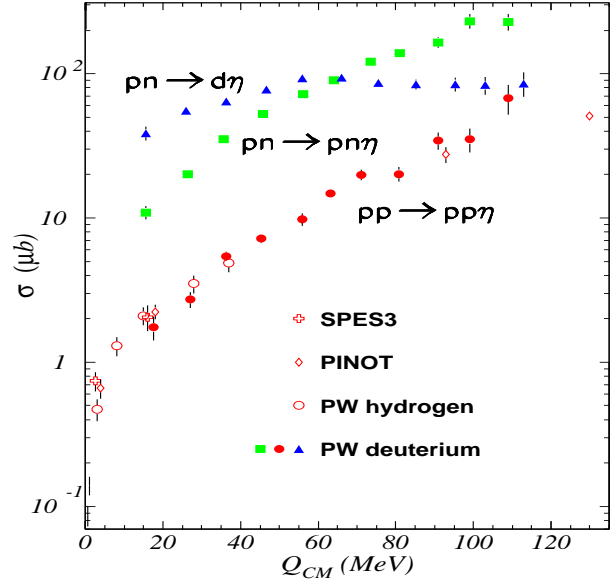}} 
\end{center}
\caption{Cross  section  for quasi-free  $\eta$ production  in $pp$  and $pn$
interactions.}\label{pnxs} 
\end{figure}

Production of $\eta$ mesons in proton-nucleon collisions has been studied using
both hydrogen and deuterium as targets.  By exploiting the neutron Fermi
momentum in the deuteron, the excitation functions for the reactions
$pn\rightarrow pn \eta$ and $pn\rightarrow d\eta$ have been measured for CM
excess energies from 15 MeV to 115 MeV (Fig.~\ref{pnxs}). The measured energy
dependence of the cross section for the $pn\rightarrow d\eta$ reaction is the
first firm evidence of the importance of the $S_{11}$(1535) nucleon resonance
in the reaction mechanism (\cite{CA+96b}).  The cross section for the reaction
$pn\rightarrow pn \eta$ was found to be about six times larger than that of the
reaction $pp\rightarrow pp\eta$.

Recently, the measurements of the $pn\rightarrow d \eta$ were extended down to
the threshold by using the CELSIUS bending magnets as a spectrometer to
identify deuterons at 0 degree.  The cross section shows an enhancement at the
threshold (\cite{CA+97}) which might be signal of a quasi-bound $d\eta$ state
suggested by \cite{Ueda}.

Windowless thin internal targets offer unique possibilities to detect spectator
protons.  The way for systematical studies of $p+n$ reactions at CELSIUS is
provided by placing semiconductor detectors inside the scattering chamber.  An
experiment which develops this technique has started recently.

The WASA/PROMICE setup will be dismounted and a new central detector with close
to 4$\pi$ coverage for photons and charged particles will be installed during
1998.  The $\eta$ decay studies will start and the meson production programme
will be continued.
\subsection{$\gamma\pi\to\pi\pi$}
The value of the decay in the chiral limit and at zero for all the kinematical
variables is predicted by the chiral anomaly. A precise measurement
is thus a good test of this fundamental phenomenon. The prediction
by the anomaly can be expressed in terms of a formfactor $F_{3\pi}$.
In the chiral anomaly this is given by
$F_{3\pi}= e/\left(4\pi^2 F_\pi^3\right)\approx9.7~GeV^{-3}$. This disagrees
at the two sigma level with the experimental measurement of \cite{antipov};
$F_{3\pi} = 12.9\pm0.9\pm0.5$. The chiral loop corrections have been
estimated by \cite{BBC} and are 6 to 12 \% over the phase space relevant
for this experiment. The loop parts are fairly small and the major part of
the upwards corrections comes from the tree level contributions
at order $p^6$ estimated via Vector Meson Dominance. Higher order corrections
have been estimated by \cite{Holstein} by unitarizing the CHPT calculation
of \cite{BBC}. An alternative way to estimate this is to use models
of the meson structure. The results are similar to the VMD calculations and
the unitarized calculation of \cite{Holstein}. An example of a recent
calculation in this approach is (\cite{Roberts}).

One recurring theme here is that the extrapolation used by \cite{antipov}
is not too reliable, there are large corrections from the measured region
to the $s=t=u=0$ point. We are therefore lucky to have several new
experiments that will test this in the future. There is
SELEX at Fermilab and COMPASS at CERN as discussed in the plenary
talk by \cite{Moinester}. These are Primakoff type experiments using
a pion beam scattering of the electromagnetic field of a nucleus.
The opposite method, scattering a photon beam of a hadron target and
extrapolating the result to the pion pole will be taken by
the CLAS experiment at TJNAF (\cite{Miskimen}). 
\section{Decays}
\label{decays}
\subsection{$\eta$}
The progress in $\eta$ decays in the last years has been fairly slow.
A few new results have been presented by the Crystal Barrel (\cite{Amsler})
about $\eta\to\pi\pi\pi$, these are given in \cite{Bijnens}. The other new
results are a preliminary limit on the branching ratio of
$\eta\to\pi^0\pi^0$ of
$7\times 10^{-4}$ from VEPP 2M in Novosibirsk (\cite{VEPP2M}).
There is also a new limit on $\eta\to e^+e^-$ from CLEO at CESR,
(\cite{CLEOeta}).

A list of accelerators and associated experiments that perform experiments
or will do so in the near future is in Table \ref{tableeta}.
\begin{table}
\caption{Accelerators and experiments that will study the $\eta$ meson.}
\label{tableeta}
\begin{center}
\begin{tabular}{|c|c|c|}
\hline
Accelerator & Experiment & Comments\\
\hline
AGS & Crystal Ball & neutral decays, to $\approx 10^{-6}$\\
CELSIUS & WASA & first $10^{-7}$ later to $10^{-9}$--$10^{-10}$\\
DA$\Phi$NE & KLOE & to $10^{-7}$--$10^{-8}$\\
GRAAL/ESRF & Lagrange & production and neutral decays\\
\hline
VEPP2M. Novosibirsk & CMD-2 & \\
MAMI & TAPS & production\\
ELSA & Crystal Barrel and TAPS & production\\
CEBAF & CLAS & production\\
CESR & CLEO & \\
Serpukhov & GAMS &\\
\hline
\multicolumn{3}{|c|}{\ldots}\\
\hline
\end{tabular}
\end{center}
\end{table}

One problem that still needs clarification is $\Gamma(\eta\to\gamma\gamma)$.
The old Primakoff experiment has a significantly lower value for
the width than the more recent experiments in $e^+e^-$--colliders.
This width is used to normalize all the other $\eta$ widths and is therefore
rather important.

Other $\eta$ decays are $\eta\to3\pi$ and $\eta\to\pi^0\gamma\gamma$ as
discussed in \cite{Bijnens}. In the anomalous decays,
$\eta\to\gamma^*\gamma^*$, only the two real photons
and the single Dalitz decays $\eta\to\gamma e^+e^-,\gamma\mu^+\mu^-$
have been measured with a single measurement of the formfactor in the latter.
As discussed below for the Kaon decays and in \cite{Bijnens} for the muon
anomalous magnetic moment, it is fairly important to know whether
double Vector Meson Dominance holds, or whether something else needs to
be used for the case when the two photons are off-shell. Measurements
of the double Dalitz decays and the formfactors in there as well as
double tagged production experiments of $\pi^0,\eta$ and $\eta'$  are therefore
very welcome.
The decay $\eta\to\pi^+\pi^-\gamma$ has a width of $58\pm6~eV$ while
the one-loop CHPT result is about 47~$eV$ (\cite{BBC}).
Improved measurements of the form-factors and the decay with
the photon off-shell would provide information about precisely which type of
contributions are the reason for this discrepancy.

On the theoretical front there have been several studies of how to
include $\eta-\eta'$ mixing in CHPT. In \cite{GL} the massive degree
was integrated out and treated perturbatively, more recent studies
try to include it as a propagating degree of freedom. The 
combined $1/N_c$ and chiral expansion
then provides a proper framework to discuss these questions,
(\cite{leutwyler97,Taronetal,georgi,derafaelperis,peris}).
\subsection{K}
\subsubsection{Experiments : Rare decays and CP-violation}
In the past year several new Kaon Decay modes have been observed.
The most interesting one is probably $K^+\to\pi^+\nu\bar\nu$ 
(\cite{adler1997c}) that in the long run will provide an accurate 
measurement of $V_{td}$. In the field of tests of CHPT we have seen 
measurements of $K^+\to\pi^+\gamma\gamma$ (\cite{TAK96})
and $K^+\to\pi^+\mu^+\mu^-$ (\cite{E78712}). Moreover, we 
have about 500  candidates  for $K_L\to\pi^+\pi^-e^+e^-$  
(\cite{klpipigg}) that can be useful for CP-violation studies.

In the near future we expect a significant improvement in the
measurements of $\varepsilon'/\varepsilon$. 
This constitutes our major CP-violation effort
in the $K$ system at present. We also expect 
improvements in the modes $K_L\to\gamma\gamma, \gamma\gamma^*,
\gamma^*\gamma^*$ very interesting both in the framework of CHPT and 
to estimate the `long-distance contamination' of more rare modes, 
as discussed below.

In the longer term we expect to observe direct CP violation in
$K_L\to\pi^0e^+e^-$ and $K_L\to\pi^0\nu\bar{\nu}$. In these channels,
as well as in $K^+\to\pi^+\nu\bar\nu$, the progress in rare kaon decays
is complementary and competitive with the one in the $B$-system.
\subsubsection{Experiments : $K_{l4}$}
Here there is the BNL experiment as presented by J.~Lowe at this
meeting (\cite{Lowe}) and the progress to be expected from the KLOE
experiment at DA$\Phi$NE. 
Both of them aim at about 300000 events in a much better acceptance environment
than the previous experiment (\cite{Rosselet}). The 4 form factors should
all be measurable. The $F$ and $G$ form factor provides tests of $\pi\pi$
scattering (\cite{Ecker}) but their absolute value is important in
determining the values of the input parameters for CHPT. KLOE should at the
same time be able to measure the decays $K^+\to\pi^0\pi^0 e^+\nu$
and $K^0\to\pi^0\pi^- e^+\nu$ as well. These latter only depend on one
form-factor to a much larger extent thus allowing for an independent
cross-check of the decay with two charged pions. At the level of statistics
expected this will also provide an important check of the isospin
breaking corrections that might be needed at the expected level of
prediction.
Once $F$ and $G$ are well measured we can turn to the anomalous formfactor $H$
where the pattern of $SU(3)$ breaking corrections is expected to be quite
different from the one in $F$ and $G$. And, last but not least, once $F$ and
$G$ are accurately determined in the electronic decay modes, we can then
use these to determine $R$ in the muonic decay mode. That way we
can determine the otherwise elusive parameter $L_4$ of the CHPT
Lagrangian and provide additional tests of the $1/N_c$ approximations.
\subsubsection{Theory: $K\to\pi\pi,\pi\pi\pi$}
The starting tool for a theoretical analysis of non--leptonic kaon 
decays is the chiral realization of the effective four--quark 
Hamiltonian for $|\Delta S|=1$ transitions.  
At the leading order ($O(p^2e^0)$ + $O(p^0e^2)$) there are only 
three independent  operators (and thus three unknown couplings)
transforming under $SU(3)_L\times SU(3)_R$ as the effective 
four-quark Hamiltonian, i.e. as $(8_L,1_R)$, $(27_L,1_R)$ and $(8_L,8_R)$. 
As shown by \cite{KMW89},
the number of independent couplings increases substantially 
at next--to--leading order: limiting to the presumably dominant 
$(8_L,1_R)$ sector there are  already 37 independent
operators (\cite{EKW}). Nonetheless, the situation is not 
dramatic since there exist various classes of processes where 
only few couplings appear and the predictive power of the theory is
not destroyed (see e.g. \cite{deRafael,Eckerrep,DEIN95,DI96}, 
and references therein).

$K\to\pi\pi$ amplitudes are typically used as input to
determine the low--energy constants of the weak chiral 
Lagrangian. Even though some recent progress has been 
made in understanding the origin of the $\Delta I=1/2$ rule
(see e.g. \cite{Antonelli} and references therein) a reliable 
prediction of $K\to\pi\pi$ amplitudes from first principles
is still far. 

$K\to 3\pi$ decays is one of the classes where the chiral
Lagrangian approach leads to interesting predictions even 
at next--to--leading order. In particular, the quadratic 
slopes in the Dalitz variables are unambiguously predicted 
in terms of widths and linear slopes (\cite{KMW91,KDHMW}). The 
predictions of $\Delta I=1/2$  quadratic slopes
are in good agreement with present data. On the other hand,
the situation in the $\Delta I=3/2$ sector is still 
unclear because of large experimental error. The situation will
certainly improve in the next years with the advent of new
high precision data. However, it must be stressed that also 
a complete theoretical analysis of the isospin--breaking terms 
is needed in order to make a precise test in the 
$\Delta I=3/2$ sector. 

Another interesting aspect of $K\to 3\pi$ decays is the 
interference measurement of $K_{L,S}\to \pi^+\pi^-\pi^0$,
that could lead to determine both the suppressed 
$K_{S}\to \pi^+\pi^-\pi^0$ amplitude and the $3\pi$ rescattering 
phases (see e.g. \cite{DIPP} and references therein). 
Here the chiral predictions are again quite firm, thus interesting 
consistency tests can be made. Moreover,
a determination of $3\pi$ rescattering phases is important to 
estimate the magnitude of direct $CP$ violation in $K\to 3\pi$. 
The interference effect in $K_{L,S}\to \pi^+\pi^-\pi^0$
has been recently observed by \cite{CPLEAR} and \cite{Zou},
showing a clear signal for a non--vanishing 
$K_{S}\to \pi^+\pi^-\pi^0$ amplitude and a preliminary 
evidence for the $3\pi$ phases. Substantial
improvements are expected at DA$\Phi$NE.

Strictly related to $K\to 3\pi$ are also
$K\to 3\pi\gamma$ transitions. These have been
completely analyzed at next--to--leading order by
\cite{DEIN96} and could offer interesting
tests of the theory. Unfortunately
the present experimental information about these
transitions is very poor, but again substantial
improvements are expected in the near future.
\subsubsection{Theory: Rare decays}
Radiative non--leptonic kaon decays may play a crucial role in our 
understanding of fundamental questions like the validity of the 
Standard Model, the origin of CP violation  
and the realization of chiral symmetry in the 
framework of non--leptonic weak interactions 
(see \cite{DEIN95,DI96}, and references therein).
For instance the measurement of $K_L \rightarrow  \pi^0 \nu \nu $ 
and $K^\pm \rightarrow  \pi^\pm \nu \nu $ rates
should allow a determination of the CKM matrix elements
competitive and complementary to the one 
attainable from $B$ decays (\cite{litten89,BuBu96}). 

In the following we will discuss some  
recent theoretical progress in various 
channels.

\underline{{\it \kpiggtot} and {\it \kggs}.}
The amplitude for 
$K_L(p) \rightarrow \pi^0 \gamma(q_1) \gamma(q_2)$
can be generally decomposed 
in terms of two independent Lorentz and gauge invariant 
amplitudes: $A(z,y)$ and $B(z,y)$, where 
$y  = p \cdot (q_1 - q_2)/m_K^2 $ and  
$ z \, = \, (q_1 + q_2)^2/m_K^2 $.
Then the double differential rate  is given by 
\begin{equation}
\Frac{\partial^2 \Gamma}{\partial y \, \partial z}   = 
\, \Frac{m_K}{2^9 \pi^3} \left[ \, z^2 \, | \, A \, + \, B \, |^2 \,
 + \,  \left( y^2  -  \Frac{\lambda (1,r_{\pi}^2,z)}{4} \right)^2
 \, | \, B \, |^2 \, \right]~, 
\label{eq:doudif} 
\end{equation}
where $\lambda (a,b,c)$ is the usual kinematical function and 
$r_{\pi} =  m_{\pi}/m_K $. Thus in the 
region of small $z$ (collinear photons) the $B$ amplitude is dominant
and can be determined separately from the $A$ amplitude. This feature
is important in order to evaluate the CP--conserving contribution
$K_L \rightarrow \pi^0 \gamma \gamma \rightarrow \pi^0 e^+ e^-$. Both
on--shell and off--shell two--photon intermediate states generate, through
the $A$ amplitude, a contribution to $K_L \rightarrow \pi^0 e^+ e^-$
that is helicity suppressed (\cite{EPR88}). Instead
the $B$--type amplitude, though appearing only at \ops, generates 
a relevant unsuppressed contribution to 
$K_L \rightarrow \pi^0 e^+ e^-$ through the 
on--shell photons, due to the different 
helicity structure.

The leading finite \opc amplitudes of \kpiggn were evaluated
by \cite{DE87a,DE87b}, and 
generate only the $A$--type amplitude in Eq.~(\ref{eq:doudif}).
The observed branching ratio for \kpiggn 
is $(1.7 \pm 0.3) \times 10^{-6}$ (\cite{PDG}),
about 3 times the \opc prediction.
However, the \opc spectrum of the diphoton invariant mass nearly agrees with
the experiment, in particular no events for small $m_{\gamma \gamma}$ are
observed, implying a small $B$--type amplitude. Thus \ops\ corrections have
to be important. No complete calculation is available, but the 
supposedly larger contributions are known~: \ops\ unitarity 
corrections (\cite{CD93,CE93,KH94}) enhance the 
\opc branching ratio by $30 \%$,
and generate a $B$--type amplitude. \cite{EP90} 
parameterized the \ops\ vector meson exchange contributions
by an effective vector coupling $a_V$.
There are two sources for $a_V$~:
i) strong vector resonance exchange with an external weak transition 
($a_V^{ext}$), and
ii) direct vector resonance exchange between a weak and a strong 
$VP\gamma$ vertices ($a_V^{dir}$). Then
$
a_V \, = \, a_V^{ext} \, + \, a_V^{dir} 
$.
The first one is model independent and gives $a_V^{ext} \simeq 0.32$ 
(\cite{EP90}), while the direct contribution depends strongly on the model
for the weak $VP\gamma$ vertex. (\cite{CE93}) noticed that one could,
simultaneously, obtain the experimental spectrum and width of 
\kpiggn with $a_V \simeq -0.9$. 
The question of the relevant \ops\ contributions relative
to the leading \opc result for \kpiggp, (\cite{EPR88}), 
can also be studied.
Unitarity corrections, (\cite{DP96}),
generate a $B$--type amplitude
and increase the rate by a  $30$--$40 \%$ while, differently from \kpiggn,
vector meson exchange is negligible in \kpiggp.
The detection by \cite{TAK96} of a few
 events,
confirms the relevance of the unitarity corrections
and a first measurement (though with large error)
 of the \opc local contributions.

The decay $K_L \rightarrow \gamma (q_1, \epsilon_1) 
\gamma^* (q_2, \epsilon_2)$  is given by an amplitude $A_{\gamma 
\gamma^*}(q_2^2)$ that can be expressed as 
$
A_{\gamma \gamma^*} (q_2^2)   =     A_{\gamma \gamma}^{exp} f(x)  $,
where $A_{\gamma \gamma}^{exp}$ is the experimental 
$A(K_L \rightarrow \gamma \gamma)$  amplitude
and $x=q_2^2/m_K^2$. The form factor
$f(x)$ is normalized to $f(0)=1$ 
and the slope $b$  is defined as 
$ f(x)  =  1   +   b x  $. 
Traditionally  experiments 
(see \cite{PDG}) do not measure directly the slope but 
they input the  full form factor suggested by 
\cite{BMS83}. However, as discussed by \cite{DPN}, 
it is more appropriate to
measure directly the slope, they estimate  
$b_{exp} \, = \, 0.81 \pm 0.18 \,$.

\cite{DPN} have analyzed the general effective weak coupling
$VP\gamma$ contributing at \ops\ to both \kpiggtot and \kggs.
The corresponding Lagrangian can be written
in terms of the 5 possible 
relevant $P\gamma$ structures and contains thus 5 
coupling constants, $\kappa_i$
to be determined from phenomenology or theoretical
models.
The Factorization Model (FM), motivated by the $1/N_c$ expansion 
(see e.g. \cite{PI91}), assumes that the dominant contribution to 
the four--quark operators of the $\Delta S=1$ Hamiltonian has
a {\em current$\times$ current} structure.
This assumption is then implemented with a 
bosonisation of the left--handed quark currents.
Applying the factorization procedure to the
construction of the weak $VP\gamma$ vertex and 
integrating out the vector mesons afterwards, (\cite{DPN}) 
have identified {\it new contributions} to the left--handed 
currents and therefore to the chiral structure 
of the weak amplitudes at \ops. Within this prescription, called
Factorization Model in the Vector couplings (FMV),
one can determine the  $\kappa_i$
and then  predict  both the $a_V^{dir}$ parameter of  \kpiggn and 
the slope $b_D$ of \kggs. The results are
$a_{V}^{dir} \, \left|_{FMV}  \, \simeq \, -0.95 \right.$  
and  $b_D^{octet} \; \left|_{FMV} \, \simeq \, 0.51 \right.$,
leading to  (see Ref. \cite{DPN} for a thorough discussion)
$a_V \, \simeq  \, -0.72$ and $b \,  \simeq  \, 0.8 - 0.9$, 
in good agreement with  phenomenology. 

\underline{\it $K_L \rightarrow \mu^+ \mu^-$.}
To fully exploit the potential of \klmm in probing short--distance 
dynamics it is necessary to have a reliable control on its long--distance
amplitude. However the dispersive contribution generated by the 
two--photon intermediate state cannot be calculated in a model independent
way and it is subject to various uncertainties. The branching
ratio can be generally decomposed as 
$
B(K_L \rightarrow \mu^+ \mu^-) \, = \, |\RcA|^2 \, + \, |\IcA|^2
$,
and the dispersive contribution can be rewritten as 
$\RcA \, = \, \RcA_{long} \, + \, \RcA_{short}$.
Within the Standard Model $\RcA_{short}$
has been evaluated at NLO by \cite{BuBu}
and is proportional to $(1.3 - \rho)$, 
where $\rho$ is the usual CKM parameter in the Wolfenstein convention.
The measurement of $B(K_L \rightarrow \mu^+ \mu^-)$ by
\cite{BNL} is almost saturated by the absorptive amplitude leaving
very little room for the dispersive contribution~:
$
| \RcA_{exp} |^2 \, = \, (-1.0 \pm 3.7) \times 10^{-10}
$ or
$| \RcA_{exp} |^2 \, < \, 5.6 \times 10^{-10} $ at $90 \%$ C.L.
Thus an upper bound on $|\RcA_{long}|$ can be used to
set a lower bound on $\rho$.

To estimate  $\RcA_{long}$, \cite{DIP} proposed
a low energy parameterization of the $K_L \rightarrow \gamma^* \gamma^*$
form factor that include the poles of the lowest vector meson 
resonances with arbitrary residues
\begin{equation}
f(q_1^2,q_2^2)  =  1 + \alpha \left(
F(q_1^2)+F(q_2^2) \right)   +  \beta  F(q_1^2)F(q_2^2)\,;
F(q^2)= q^2/(q^2-m_V^2)
\label{eq:fq1q2} 
\end{equation}
$\alpha$ and $\beta$, expected to be ${\cal O}(1)$
by naive dimensional chiral power counting, are in principle directly
accessible by experiment in \kgll and $K_L \rightarrow 
e^+ e^- \mu^+ \mu^-$. Up to now there is no experimental 
information on $\beta$, whereas  $\alpha = -1.63 \pm 0.22$. Note 
that an accurate theoretical determination of $\beta$
requires the knowledge of the strong Pseudoscalar-Vector-Vector vertex.

The form factor defined in Eq.~(\ref{eq:fq1q2}) goes as $1 + 2 \alpha + 
\beta$ for $q_i^2 \gg m_V^2$ and, as long as $1 + 2 \alpha + \beta\not=0$,
an ultraviolet cutoff is needed
to regularize the contribution to $\RcA_{long}$.
However, for $q_i^2 \gg m_V^2$ one can use perturbative QCD 
to estimate the  $K_L \rightarrow \gamma^* \gamma^*$ vertex. 
The explicit QCD calculation 
show a mild behavior of the form factor at large $q^2$, 
consistent with result of the FMV model. 
Using $\alpha_{exp}$ and the QCD constraint \cite{DIP}
estimate
\begin{equation} 
\rho > -0.42 \; \; \; (90 \% C.L.)~.
\label{eq:rhobo}
\end{equation}
This bound could be very much improved if $\alpha$, $\beta$
and possibly higher order contributions to the 
$K_L \rightarrow \gamma^* \gamma^*$ form factor
were measured more accurately.

\underline{$K_L \rightarrow \pi^0 e^+ e^-$.}
This process is being searched for as a signal of direct $\Delta S =
1$ CP violation. We analyze the three components of the decay: 1)
direct CP violation through one-photon exchange, 2) CP violation
through the mass matrix and 3) CP-conserving (two-photon)
contributions. The primary weak Hamiltonian responsible for the
transition has the form
\begin{equation}
{\cal H}^{\Delta S = 1}_W = {G_F \over \sqrt{2}}
\left[ C_{7V} (\mu) Q_{7V} + C_{7A} Q_{7A} \right],
\end{equation}
where $Q_{7V} =  ( \ov{s} d)_{V-A} (\ov{e} e)_V$ and
$Q_{7A} = ( \ov{s} d)_{V-A} (\ov{e} e)_A$.
The decay rate is
\be
B(K_L \rightarrow \pi^0 e^+ e^- )_{\rm dir} = 4.16 ({\Im m} \lambda_t )^2
(y^2_{7A} + y^2_{7V}),
\ee
${\Im m} \lambda_t = {\Im m} V^{\phantom{*}}_{td}
V^*_{ts} = \vert V^{\phantom{*}}_{ub} \vert \vert V^{\phantom{*}}_{cb} \vert
\sin \delta = A^2 \lambda^5 \eta$,
and $V^{\phantom{*}}_{td} = \vert V^{\phantom{*}}_{ub}
\vert \sin \delta$. $A, \lambda,
\eta$ refer to the Wolfenstein parametrization of the CKM matrix.
Using the results of \cite{Buras} for $y_{7A}$ and
$y_{7V}$, one gets the branching ratio
\begin{equation}
\label{21}
B (K_L \rightarrow \pi^0 e^+ e^-)_{\rm dir} = 2.32 \times 10^{-12}.
\end{equation}

For the second component, we review the formalism for analyzing mass
matrix CP violation, first set forth by
\cite{EPR1} and \cite{EPR88}. This amounts to
a prediction for the decay rate for $K_S \rightarrow \pi^0 e^+ e^-$, since
the mass matrix effect is defined by
\begin{eqnarray}
\label{22}
A (K_L \rightarrow \pi^0 e^+ e^-) \vert_{\rm MM} & \equiv & \epsilon A
(K_S \rightarrow \pi^0 e^+ e^-).
\end{eqnarray}
The uncertainties in this method appear to be so large that they will
obscure the direct CP violation unless it is possible to measure the
$K_S\rightarrow \pi^0 e^+ e^-$ rate, which could become possible at
DA$\Phi$NE. In particular, one cannot extract all the parameters
needed from the available experimental data. Using
additional assumptions instead, one gets for the mass-matrix contribution
\begin{equation}
\label{43}
B ( K_L \rightarrow \pi^0 e^+ e^-)_{\rm MM} = 0.37 \times 10^{-12}.
\end{equation}

The CP-conserving amplitude remains also
somewhat uncertain, but pre\-sent indications are such that there may be a
sizable CP-violating asymmetry in the $e^+, e^-$ energies from the
interference of CP-conserving and CP-violating amplitudes. This may
potentially be useful in determining whether direct CP violation is
present.
This component proceeds through
a CP-conserving two-photon intermediate state. If we ignore the electron
mass, the form of the amplitude will be
\begin{equation}
\label{50}
A(K_L \rightarrow \pi^0 e^+e^-)_{\rm CPC} =
G_8 \alpha^2 K p^{\phantom{l}}_K \cdot (k - k^{\prime})
 (p^{\phantom{l}}_K +
p^{\phantom{l}}_{\pi})^{\mu}
\ov{u} \gamma_{\mu} v,
\end{equation}
where $K$ is given by (\cite{DG95})
\begin{equation}
K = {B(x) \over {16 \pi^2 m^2_K}} \left [
{2 \over 3} \ln \left({m^2_{\rho}} \over {-s}\right)
- {1 \over 4} \ln \left({-s} \over {m^2_e}\right)
+ {7 \over {18}}
\right ],
\end{equation}
and $B$ is the only form factor relevant in the limit of vanishing
$m_e$. Using the estimate of  $B$ by \cite{CE93} one gets
\begin{equation}
B(K_L \rightarrow \pi^0 e^+ e^-)_{\rm CPC} = 4.89 \times 10^{-12}.
\end{equation}
A slightly smaller result is obtained in (\cite{DPN})
due to the inclusion of the corrections suggested by \cite{KH94}:
\begin{equation}
0.1 < B(K_L \rightarrow \pi^0 e^+ e^-)_{\rm CPC}\times 10^{-12} < 3.6. 
\end{equation}

\underline{{\it $K_S \rightarrow \pi^0 e^+ e^-$}  and 
{\it $K^\pm \rightarrow \pi^\pm l^+ l^-$}.} Contrary to 
$K_L \rightarrow \pi^0 e^+ e^-$, these processes are dominated by 
long--distance effects. The amplitudes can be calculated using the
parametrization of \cite{EPR1,EPR88}:
\begin{eqnarray}
A(K_S \rightarrow \pi^0 e^+ e^-) &=& - {G_8 \alpha \over 4 \pi}
d_S (p^{\phantom{l}}_K + p^{\phantom{l}}_{\pi})^{\mu} \ov{u}
\gamma_{\mu} v   \\
A(K^+ \rightarrow \pi^+ l^+ l^-) &=& - {G_8 \alpha \over 4 \pi}
d_+ \left[ (p^{\phantom{l}}_K + p^{\phantom{l}}_{\pi})^{\mu} 
-\frac{(m_K^2-m_\pi^2)}{q^2}q^{\mu} \right] \ov{u}\gamma_{\mu} v
\nonumber 
\end{eqnarray}
with
\begin{eqnarray}
\label{37}
d_S & \equiv & {\Re e} \; w^{\phantom{2}}_S + 2 \phi_K (q^2), \qquad
d_+ \equiv  w^{\phantom{2}}_+ +  \phi_K (q^2)
+  \phi_\pi (q^2), \nonumber \\
w_+ & = & {64\pi^2 \over 3} \left( N_{41}^r -N_{15}^r + 3 L^r_9
\right) - {1 \over 3} \ln {m^2_K \over \mu^2 };\, 
w^{\phantom{2}}_S  =  w_+ + 32\pi^2 \left( N_{15}^r - 2 L^r_9
\right), \nonumber \\
\phi_i (q^2) & = & {m^2_i \over q^2} \int^1_0 dx \left[ 1 - {q^2 \over
m^2_i} x (1 - x) \right] \ln \left[ 1 - {q^2 \over m^2_i} x (1 - x) \right],
\end{eqnarray}
where the $N^r_i$'s are the renormalized couplings of the 
\opc $\Delta S$ = 1 weak Lagrangian in the basis of 
\cite{EKW}. $w_+$ is extracted from rate and/or the spectrum 
of  $K^\pm \rightarrow \pi^\pm l^+ l^-$, however  
recently \cite{E78712} seems to confirm  a value 
of $w_+$ in $K^+ \rightarrow \pi^+ \mu^+ \mu^-$
2$\sigma$'s away from the value extracted  from 
$K^+ \rightarrow \pi^+ e^+ e^-$.

\underline{\it $K_L \rightarrow \pi^0 \gamma e^+ e^-$}
This decay, which occurs at a {\it higher} rate than the nonradiative
process $K_L \rightarrow \pi^0 e^+ e^-$, can be a background to CP
violation studies using the reaction in the previous section. It is of interest
in its own right in the context of CHPT,
through its relation to the decay $K_L \rightarrow \pi^0
\gamma\gamma$. Using the framework of the
calculation performed by \cite{CE93} for $K_L \rightarrow
\pi^0 \gamma\gamma$, one can provide a straightforward
\ops\ calculation.
This is the generalization to
$k^2_1 \neq 0$ of the
original chiral calculation of \cite{CD93,CE93}. Here $k_1$ is the
momentum of the off-shell photon. The
branching ratio obtained by \cite{DG97} is:
\begin{equation}
{\rm B}(K_L \rightarrow \pi^0 \gamma e^+e^-) = 2.3 \times 10^{-8}.
\end{equation}

The behavior of the $K_L \rightarrow \pi^0 \gamma e^+ e^-$ amplitude
mirrors closely that of the process $K_L \rightarrow \pi^0 \gamma
\gamma$.
This reaction should be reasonably amenable to experimental
investigation in the future. It is 3--4 orders of magnitude larger
than the reaction $K_L \rightarrow \pi^0 e^+ e^-$, which is one of the
targets of experimental kaon decay programs, due to the connections of
the latter reaction to CP studies. The regions of the distributions where the
experiment misses the photon of the radiative process can potentially
be confused with $K_L \rightarrow \pi^0 e^+ e^-$ if the resolution is
not sufficiently precise. In addition, since the $\pi^0$ is detected
through its decay to two photons, there is potential confusion related
to misidentifying photons. The study of the reaction $K_L \rightarrow
\pi^0 \gamma e^+ e^-$ will be a valuable preliminary to the ultimate
CP tests.
\subsection{$\pi$}
For $\pi^0\to\gamma\gamma$ we discussed the influence of the anomaly,
CHPT calculations for this process and the ones with one or two off-shell
photons, Dalitz or double Dalitz decays exist up to order $p^6$
(\cite{anomaly}). The corrections are mainly the change of $F_0$, the
chiral limit decay constant, to $F_\pi$ and a small contribution to the slope.
The main contributions are those coming from the $p^6$ Lagrangian.
VMD, the chiral quark model and the Nambu--Jona-Lasinio model (\cite{BP})
have all been
used to estimate these constants. The slope agrees reasonably well in
these three estimates. A remeasurement of this slope would be useful. The
present measurements rely on an extrapolation from high $Q^2 > 1~GeV^2$
to small values. The main uncertainties on the theoretical prediction are
now the value of $F_{\pi^0}$ where we have to remove isospin breaking and
electromagnetic corrections from the measured $F_{\pi^+}$ and the
quark mass corrections to the decay rate. The latter are large in the
chiral quark model but unlike most other cases the ENJL model predicts
in this case a vastly smaller correction of about $+0.7\%$ (\cite{BP}).
Notice that both of these corrections tend to increase the decay width.

The other decay we discussed was the recent two-loop calculation
of $\pi\to e\nu\gamma$ (\cite{BT}). The results of this work were presented
in (\cite{Bijnens}).
\section{Production II}
\label{productionII}
\subsection{$a_\mu=\frac{g_\mu-2}{2}$ and $\alpha(M_Z^2)$}
The high precision muon $g-2$ measurements are excellent
candidates for probing the electro-weak sector of the Standard Model (SM)
and new physics scenarios beyond the Standard Model. The present experimental
value is \cite{CERN}:
\be
a_\mu = (g_\mu-2)/2 = 11\, 659\, 230(85)\times 10^{-10}\,
\mbox{($\mu^\pm$ average)} .
\label{EXP}
\ee
The muon $g-2$ experiment E821 at BNL (\cite{BNLm}) plans to
reach $\pm 40\times 10^{-11}$.
This is sufficient to observe the electro-weak 
contributions to $a_\mu$ if the precision of the hadronic
corrections is improved. The SM prediction of $a_\mu$ can be
written as
$a_\mu = a_\mu^{QED}+a_\mu^{hadronic}+a_\mu^{e.w.}$.
The dominant pure QED contribution to $a_\mu$ is known 
up to $O(\alpha /\pi )^5$ (\cite{QED}).
SM electro-weak contributions are known up to two loops (see \cite{weak} and
refs. therein). Numerical values are in Table \ref{mutab2}.

The three classes of dominant hadronic contributions are: (I)
hadronic
vacuum polarization (HVP) at order $(\alpha /\pi)^2$, (II) higher
order corrections to HVP ($(\alpha /\pi)^3$) and (III) light-by-light
(LL) ($(\alpha /\pi)^3$). (II) has been numerically
computed in \cite{KNO} with the value $-90(5)\times 10^{-11}$, while a recent
analytical estimate (\cite{Krause}) gives $-101(6)\times 10^{-11}$.
We focus here on some recent developments in the determination of (I) and
open questions in (III), further details on the latter are given by
\cite{Bijnens}. The most precise determination of the
HVP contribution to $a_\mu$ (see \cite{Rafael,Pallante} for an alternative
theoretical estimate) and $\alpha^{-1} (M_Z^2)$ is extracted from
$R(s)=\sigma(e^+e^-\to\mbox{hadrons})/\sigma(e^+e^-\to\mu^+\mu^-)$
through the following dispersion 
relations ($\Delta \alpha_{had}^{(5)}$ is for five light quarks (u,d,s,c,b)):
\be
\left[a_\mu^{HVP};\Delta \alpha_{had}^{(5)} (M_Z^2)\right]=
\left[
\left ({\alpha m_\mu\over 3\pi}\right )^2;
 -{\alpha M_Z^2\over 3\pi}\right]
\int_{4m_\pi^2}^\infty
ds~ {R(s)\left[\hat K(s);1\right] \over
\left[ s^2 ; s(s-M_Z^2-i\epsilon )\right]}\, .
\label{DISP}
\ee
$\hat{K}(s)$ smoothly increases from 0.63 to 1 with $s$.
$a_\mu^{HVP}$ is dominated by the low energy $2\pi ,\rho$ region, 
while
$\Delta \alpha_{had}^{(5)}$ is dominated mainly by
$\sqrt{s}=[2, 40]$ GeV. \cite{Alemany} use
recent ALEPH data on hadronic $\tau$ decays
(\cite{ALEPH}) ($2\pi$ and $4\pi$ channels) and the $R(s)$ data
to improve the determination of
$a_\mu^{HVP}$ and $\Delta \alpha_{had}^{(5)}$.
Hadronic $\tau$ decay data in the $\pi^0\pi^-$
channel agree well with the
corresponding $e^+e^-\to (\pi^+\pi^-)^{I=1}$ data (and are more accurate)
and with CHPT predictions
near threshold (\cite{ALEPH,Alemany}). 

The largest 
uncertainties for $a_\mu^{HVP}$(\cite{Alemany})
still come from the $\rho$-meson region followed by
the $\sqrt{s}=[2.125,40]$ GeV region. 
The third 
comes from the $4\pi$ and the $KK\pi\pi$ 
channels, the former mainly because of
experimental discrepancies in $e^+e^-\to
\pi^+\pi^- 2\pi^0$.
The inclusion of $\tau$ data
highly improves the precision in 
$a_\mu^{HVP}$, which is dominated by
the $2\pi$ channel. 
The value of (\cite{Alemany}) is compatible with most previous
estimates. We therefore quote their value
for $a_\mu^{HVP}$ in Table \ref{mutab2}.
The inclusion of $\tau$ data does not improve the
determination of $\Delta \alpha_{had}^{(5)}$, since
this is dominated by the $2-40$ GeV region.
The $\tau$ data slightly 
increase previous values
from $e^+e^-$ data only, leading to
$\alpha^{-1}(M_Z^2) = 128.878(85)$(\cite{Alemany}).

The main uncertainties in the LL contribution (\cite{BPP,HKS})
come from the lack
of knowledge of the off-shell behaviour of the vertices $P\gamma^*\gamma^*$
and $P'P'\gamma^*\gamma^*$, where $P=\pi^0 , \eta , \eta^\prime$
and $P'=\pi^\pm,K^\pm$.
Both are responsible for about half the error of \cite{BPP}.
The present
discrepancy between the estimate in (\cite{BPP}) $-92(32)\times 10^{-11}$ and
the one in (\cite{HKS}) $-79.2(15.4)\times 10^{-11}$ (see \cite{Bijnens} for a
discussion) comes mainly from the model dependence of the
$P'P'\gamma^*\gamma^*$ vertex in the meson-loop diagram.
Recently, CLEO published data for the meson-photon transition form factors
with one photon highly off-shell $(\gamma^*(Q^2)\gamma\to P)$ 
and for space-like photons with
$Q^2$ from 1.5 to 9 $(\pi^0)$, 20 $(\eta )$ and 30 $(\eta^\prime )$
GeV${}^2$ (\cite{CLEO}).
Their data  support a pole dominance picture and
an asymptotic behaviour $A/Q^2$ 
suggested by theory (\cite{ASY}) but not with the suggested coefficient $A$.
In addition, the form factor 
with two off-shell photons still needs to be studied theoretically
and experimentally. 

Adding together the SM contributions, we obtain
$a_\mu$ as shown in table \ref{mutab2}: 
\begin{table}
\caption[]{Updated SM prediction of  
 $a_\mu$. Errors added in quadrature}
\label{mutab2}
\begin{center}
\begin{tabular}{lcc}
\hline
Type  & $a_\mu^{SM} (\times 10^{11})$ & Reference \\
\hline
QED & $116\, 584\, 705.7(1.9)$ & \cite{QED}\\
Weak & 151(4) & \cite{weak}\\
Hadronic-HVP & 7011(94)& \cite{Alemany} \\
Hadronic-h.o.&-90(5) & \cite{KNO} \\
Hadronic-LL &-92(32) & \cite{BPP}\\
\hline
Total & 116\, 591\, 686(100) & \\
\hline
\end{tabular}
\end{center}
\end{table}

Muon $g-2$ experiments also allow to explore new physics scenarios beyond the
Standard Model and are able to put constraints on new physics scales which are
competitive/complementary with limits from present accelerators. From the
prediction of $a_\mu^{SM}$ of table \ref{mutab2} and the
experimental value (\ref{EXP}) the allowed window for new physics is 
$-11.0 \times 10^{-9}<\delta a_\mu < 23.3 \times 10^{-9}$.
Experimental and theoretical errors have been added in
quadrature. This already imposes stringent contraints on different
SUSY scenarios (\cite{SUSY}). 
Assuming the window for new physics from $BNL-E821$ to be $|\delta
a_\mu | < 40\times 10^{-11}$, we list in table \ref{tab3} the accessible limits
on new physics. Muon $g-2$ experiments are highly preferrable to accelerator
measurements for establishing $W^\pm$ compositeness and a value for $g_W-2$, 
they are
competitive for SUSY, various Higgs scenarios and $\mu$ compositeness, but not
competitive for exploring extra $W,Z$ bosons scenarios. 
\begin{table}
\caption[]{Limits on new physics scales with a window for
of $|\delta
a_\mu | \leq 40\times 10^{-11}$. The typical contribution to $\delta a_\mu$ is
also shown for each scenario. SUSY treated in the large
$\tan\beta$ limit.}
\label{tab3}
\begin{center}
\begin{tabular}{lccc}
\hline
Type  & $\delta a_\mu$ & Limit & Quality \\
\hline
SUSY &\hskip-1cm $\tan\beta\,{\alpha m_\mu^2/( \sin^2\theta_W m_{susy}^2)}$ &
$>$120-130 GeV & competitive \\
$W_R$ & ${m_\mu^2/m_{W_R}^2}$ & $>$ 250 GeV & $>$300-450 from 
$p\bar{p}$ \\
$Z^\prime$ & ${m_\mu^2/ m_{Z^\prime}^2}$ & $>$ O(100) GeV & $>$120-130
pres. lim. \\
Light-Higgs &-$O(10^{-3}g)$ & $>$O(300) GeV & competitive \\
Heavy Higgs &-$O(g)$ & $>$ O(500) GeV & competitive \\
$\mu$ compositness & ${m_\mu^2/ \Lambda^2}$& $>$4-5 TeV & competitive\\
Excited $\mu$ &   ${m_\mu/m_\mu^*}$ & $>$ 400 GeV &  competitive\\
${(g_W-2)/ 2}$ & & $\leq 0.02$ & dominant (LEP II$\leq 0.2$) \\
$W^{\pm}$ compositness &${m_W/ \Lambda}$ & $>$ 2 TeV & dominant\\
&${m_W^2/
  \Lambda^2} $ & $>$400 GeV & dominant\\
\hline
\end{tabular}
\end{center}
\end{table}
\subsection{$VV$ and $AA$ chiral sum rules to $p^6$}
In this section we discuss a new calculation of the isospin and hypercharge
axialvector current propagators ($\Delta_{{\rm A}33}^{\mu\nu}(q^2)$
and $\Delta_{{\rm A}88}^{\mu\nu}(q^2)$) to two loops in
$SU(3)\times SU(3)$ CHPT. (\cite{GK3,GK4})
This completes work done for the corresponding quantities involving 
vector currents, (\cite{GK1,GK2}) which was reviewed in the
plenary talk by Bijnens. (\cite{Bijnens})

The motivation to consider these quantities is twofold. First, it
becomes possible to formulate new chiral sum rules, valid to second
order in quark mass. Second, these sum rules allow one to fix certain 
coupling constants of the order $p^6$ chiral lagrangian (LEC). They are
given as integrals over moments of the spectral functions of vector 
and axialvector currents. Since both sides of these sum rules are 
physical observables, the determination of the coupling constants does
not depend on the renormalization scale. This kind of uncertainty,
inherent in the often used method of resonance saturation, is therefore
avoided. Moreover, it is possible to check the validity of the
principle of resonance saturation at order $p^6$ (\cite{GK2}) To our 
knowledge, this is the first time this goal has been achieved in the 
nonanomalous sector of ChPT. Below, we shall give an example where an
order $p^6$ LEC is seen to be {\it not} saturated by the {\it lowest}
lying resonance, but rather gets substantial contributions from 
the region above this resonance. 

Besides the application to chiral sum rules discussed here, this work 
yields a set of additional results, among which are:
i) a large number of constraints on the set of beta functions of
${\cal O}(p^6)$ counterterms,
ii) predictions for threshold behaviour of the $3\pi$, ${\bar K}K\pi$,
${\bar K}K\pi$, $\eta \pi \pi$, {\it etc} axialvector spectral
functions,
iii) an extensive analysis of the so-called `sunset' diagrams; since 
we work in chiral SU(3), the case of unequal masses has to be considered,
iv) mass corrections to the Das-Mathur-Okubo sum rule (\cite{DMO}), and
v) a complete two-loop renormalization of the masses
and decay constants of the pion and eta mesons.
This final item places the axialvector problem
at the heart of two-loop studies in $SU(3)\times SU(3)$ ChPT.

The $SU(3)$ axialvector current propagators are defined as
\begin{equation}
\Delta_{{\rm A}ab}^{\mu\nu}(q^2) \equiv i \int d^4x~ e^{iq\cdot x}~
\langle 0|T\left( A^\mu_a (x) A^\nu_b (0)\right)|0\rangle
\qquad (a,b = 1,\ldots, 8)
\label{aa1}
\end{equation}
and have spectral content
\begin{equation}
{1\over \pi} {\cal I}m~\Delta_{{\rm A}ab}^{\mu\nu}(q^2) =
(q^\mu q^\nu - q^2g^{\mu\nu}) \rho_{{\rm A}ab}^{(1)}
(q^2 ) + q^\mu q^\nu \rho_{{\rm A}ab}^{(0)} (q^2) \ ,
\label{aa2}
\end{equation}
where $\rho_{{\rm A}ab}^{(1)}$ and $\rho_{{\rm A}ab}^{(0)}$ are
the spin-one and spin-zero spectral functions.  The tensor structure
of Eq.~(\ref{aa2}) motivates the usual decomposition,
\begin{equation}
\Delta_{{\rm A}ab}^{\mu\nu}(q^2) =
(q^\mu q^\nu - q^2 g^{\mu\nu}) \Pi_{{\rm A}ab}^{(1)}
(q^2 ) + q^\mu q^\nu \Pi_{{\rm A}ab}^{(0)} (q^2) \ \ .
\label{aa2a}
\end{equation}
$\Pi_{{\rm A}ab}^{(1)}$ and $\Pi_{{\rm A}ab}^{(0)}$ are
the spin-one and spin-zero axialvector polarization
functions. We consider $a=b=3$ (isospin) and $a=b=8$ (hypercharge).

We have determined the propagator through two-loop order
\begin{eqnarray}
\Delta_{{\rm A}aa}^{\mu\nu}(q^2)&=&
(F_a^2 + {\hat \Pi}_{{\rm A}aa}^{(0)}(q^2))
g^{\mu\nu}  - {F_a^2 \over q^2 - M_a^2} q^\mu q^\nu \nonumber \\
& &+ ( 2L_{10}^{\rm r} - 4H_1^{\rm r} + {\hat \Pi}_{{\rm A}aa}^{(1)}(q^2))
 (q^\mu q^\nu - q^2 g^{\mu\nu}) \ ,
\label{two-loop}
\end{eqnarray}
where $F_a^2$, $M_a^2$ are now renormalized at two-loop level,
$L_{10}^{\rm r}$, $H_1^{\rm r}$ are ${\cal O}(p^4)$ counterterms
which appear in the one-loop analysis and
${\hat \Pi}_{\rm A}^{(0,1)}(q^2)$ are finite two-loop functions.
These latter contain contributions from the order $p^6$ counterterm 
lagrangian in terms of renormalized LEC's. Note that
there are kinematic poles at $q^2 = 0$ in both
polarization functions defined in Eq. (\ref{aa2a}), 
although the sum $\Pi_{\rm A}^{(1)}$ + $\Pi_{\rm A}^{(0)}$ 
is free of such poles. 

The derivation of chiral sum rules proceeds by first obtaining
dispersion theoretic expressions for the various polarization
functions. The numbers of necessary subtractions is obtained from the 
asymptotic behavior ($s \to \infty$) of the
spectral and polarization functions, which follows 
from perturbative QCD, (\cite{FNR,NR}). Entire sequences of chiral sum 
rules are obtained by evaluating arbitrary derivatives of such 
dispersion relations at $q^2=0$. Examples are (a=3,8)
\begin{eqnarray}
&&{1\over n!}\left[{d \over dq^2}\right]^n\left(
\Pi_{{\rm V}aa}^{(1)} - \Pi_{{\rm A}aa}^{(1+0)}
\right)(0) = \int_0^\infty ds
{(\rho_{{\rm V}aa}^{(1)} - \rho_{{\rm A}aa}^{(1+0)})(s) 
\over s^{n+1}}, \ n \ge 0 \quad 
\label{aa11} \\
&&{1\over n!}\left[{d \over dq^2}\right]^n
{\hat \Pi}_{{\rm A}aa}^{(0)}(0)=\int_0^\infty ds
{{\bar \rho}_{{\rm A}aa}^{(0)}(s) \over s^n}, \quad  n \ge 1
\label{aa11a} \\
&&{1\over (n-1)!}\left[{d \over dq^2}\right]^{n-1}
{\hat \Pi}_{{\rm A}aa}^{(1)}(0) - {1\over n!}\left[{d \over dq^2}\right]^n
{\hat \Pi}_{{\rm A}aa}^{(0)}(0)=\int_0^\infty ds
{\rho_{{\rm A}aa}^{(1)}(s) \over s^n}.\quad 
\label{aa11b}
\end{eqnarray} 
Eq. (\ref{aa11b}) is valid for $n \ge 2$ and  
${\bar \rho}_{{\rm A}aa}^{(0)}(s) \equiv
\rho_{{\rm A}aa}^{(0)}(s) - F_a^2 \delta(s - M_a^2)$.

As an application, consider the sum rule of Eq.~(\ref{aa11b}) 
with $n=2$ and isospin flavour. Explicitly (\cite{GK3})
\begin{equation}
4 \left( 2 B_{32}^{(0)} - B_{33}^{(0)}\right)(\mu)
- {0.204 + \log{M_\pi^2 \over \mu^2}
+ {5\over 4} \log{M_K^2 \over \mu^2} \over
3072 \pi^4 F_\pi^2}
= \int_0^\infty ds ~
{\rho_{{\rm A}33}^{(1)}(s) \over s^2} \ .
\label{aa13}
\end{equation}
The first term inside the parentheses on the LHS
arises from a scale-indepen\-dent two-loop contribution.
$2 B_{32}^{(0)}-B_{33}^{(0)}$ is a renormalized coupling constant
of the $O(p^6)$ counterterm lagrangian.
To estimate it, we must evaluate the right-hand-side of
Eq. (\ref{aa13}). The dominant contribution to $\rho_{{\rm A}33}(s)$
arises from the ${\rm J^{PC}}=1^{++}$ $a_1$ resonance. But there is 
also some structure at higher energy due mainly to $n_\pi\geq 5$ 
multiparticle states, and fairly rapidly thereafter asymptotic 
behaviour sets in. Employing the fits of Ref. (\cite{DG}) we obtain
$(2 B_{32}^{(0)} - B_{33}^{(0)})(M_{a_1}) \simeq 0.0044~{\rm GeV}^{-2}$
with an error bar of about 15 \%. 
The ${\cal O}(p^6)$ counterterm dominates the other terms on the
LHS of Eq.~(\ref{aa13}), showing the importance of a
full ChPT calculation as compared to a chiral-log treatment.
However, the $a_1$ resonance contributes
only $\simeq 0.0030~{\rm GeV}^{-2}$, thus 
the principle of lowest lying resonance saturation is violated at 
the level of 30 \% for this combination of $O(p^6)$ LEC's. 

A phenomenological analysis of further chiral sum rules, including 
mass corrections to the Das-Mathur-Okubo sum rule (\cite{DMO}), is under 
investigation.

\subsection{Meson-meson scattering in a nonperturbative method}
In $O(p^4)$ $\chi$PT the amplitude {\em matrix} is obtained as
$T\simeq T_2+T_4+...$

We are proposing a method where (\cite{fut}) 
\begin{equation}
T\simeq T_2\cdot[T_2-\hbox{Re}T_4-T_2\cdot \hbox{Im}G\cdot T_2]^{-1}\cdot T_2
\end{equation}
with $G$ satisfying $T_2\cdot \hbox{Im}G \cdot T_2= \hbox{Im} T_4$.

If the complete $O(p^4)$ calculations are available, it reduces to:
\begin{equation}
T\simeq T_2\cdot[T_2-T_4]^{-1}\cdot T_2
\end{equation}
which is nothing but the generalization of the Inverse Amplitude Method
(\cite{IAM})
to coupled channels. However, at present, not all the $O(p^4)$ calculations 
for the $\pi\pi, \pi K, K \bar{K}, \pi \eta$ and $K\eta$ channels 
have been performed.

Nevertheless, we can use a similar approach to the Lippmann-Schwinger 
equations (\cite{OO}), using
\begin{equation}
G=i\int\frac{d^4 q}{(2\pi)^4}
\frac{1}{[q^2-m_{i1}^2+i\epsilon][(P-q)^2-m_{i2}^2+i\epsilon]]}
\end{equation}
to approximate:
\begin{equation}
\hbox{Re}T_4\simeq T_4^{tree}+T_2\cdot \hbox{Re} G\cdot T_2 
\end{equation}
and thus we get
\begin{equation}
T\simeq T_2\cdot[T_2-T_4^{tree}-T_2\cdot G\cdot T_2]^{-1}\cdot T_2
\end{equation}

\begin{figure}
\centerline{
\hbox{
\epsfig{width=12cm,file=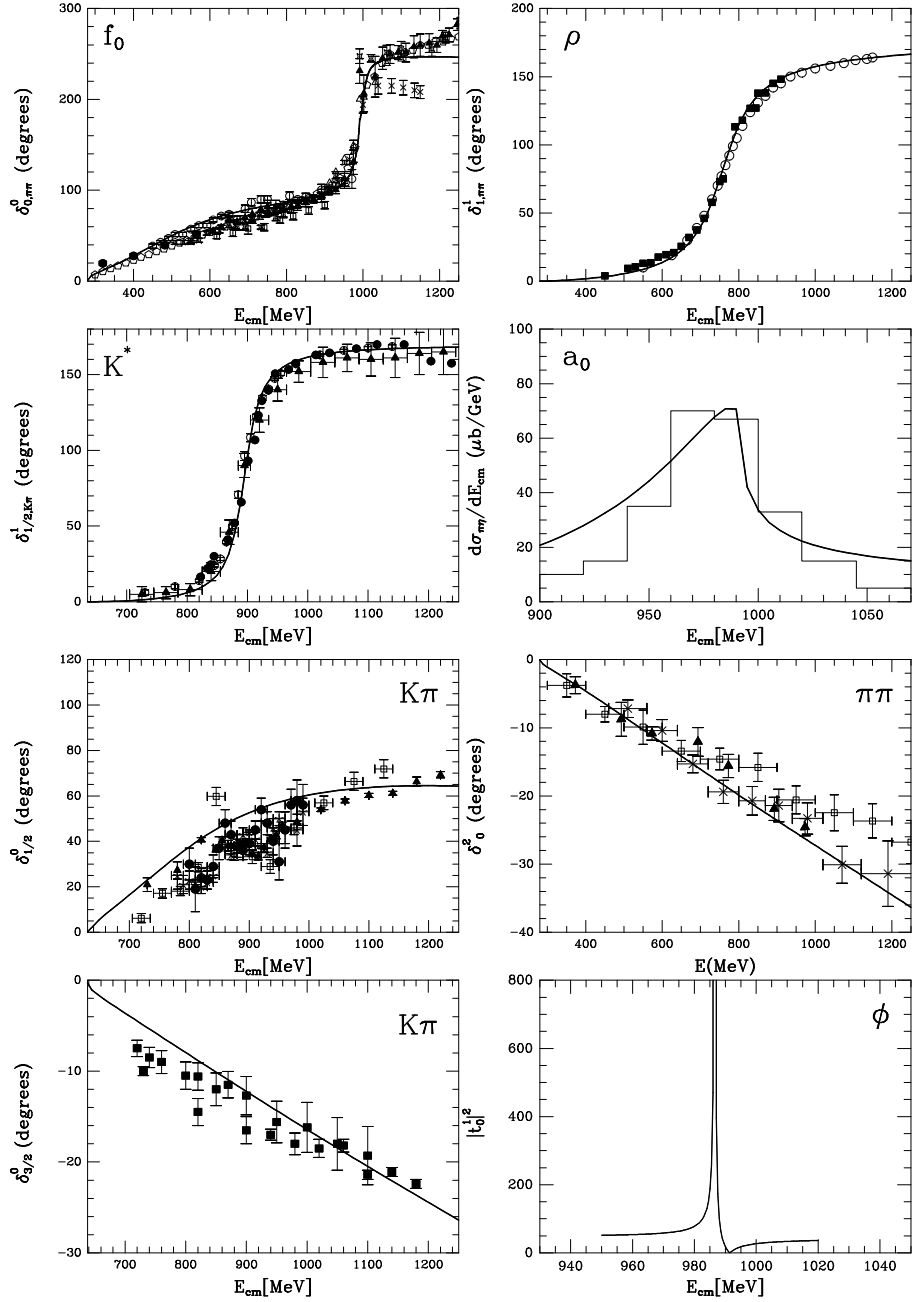}}}
\caption{The fit to the various meson scattering described in the text}
\label{figpelaez}
\end{figure}

In Fig. \ref{figpelaez} we present the results of applying 
this method to meson-meson interactions below 1.2 GeV. 
The fit has 7 free parameters (the chiral parameters in the $\chi$PT
$O(p^4)$ Lagrangian, but in a cut-off regularization scheme).
It reproduces the data (see references in (\cite{IAM}) and (\cite{OO}))
with a remarkable success up to 1.2 GeV,
including six resonances: the $\sigma,\rho,K^*,f_0,a_0$ and $\phi$.
The reasons why such an extremely simple method
works so well are still under study, together with possible
applications to other processes (\cite{fut}).
\section{Other: Divergences at order $p^6$}
\label{other}
As discussed at length in this working group, there are now
several calculations available in the literature at order $p^6$ in
CHPT. Unlike the calculations at order $p^4$, these have not
been checked against
the complete calculation of the divergences at the
level of the generating functional, since until now this was not known at
this order. 
In this talk I have described the results of this rather lengthy
calculation that we (G.~Colangelo, J.~Bijnens and
G.~Ecker\footnote{We also enjoyed the collaboration of J. Gasser in early
  stages of this project. We gratefully acknowledge his very important
  contribution. })
have just completed. 

\begin{figure}
\centerline{\epsfig{height=6cm,width=8cm,file=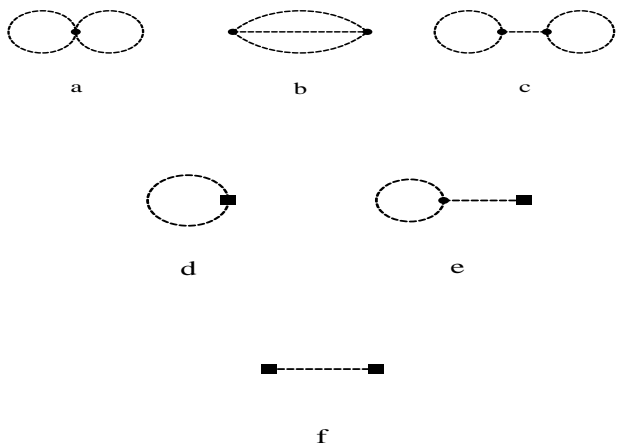}} 
\caption{Diagrams that contribute to $Z_2$. Vertices with dots stand for
  vertices coming from $S_0$ ({\em  i.e.} vertices of order $p^2$ in CHPT),
  whereas vertices with a box stand for vertices coming from $S_1$ ({\em
    i.e.} vertices of order $p^4$ in CHPT).}
\label{fig:2loop}
\end{figure}

The generating functional $Z$ is defined as:
\be
e^{{i \over \hbar } Z[f]} = {\cal N} \int {\cal D} \phi e^{ {i \over \hbar}
  S[\phi,f]} \; \; , 
\ee
where ${\cal N}$ by definition ensures $Z[0]=0$, and
$S[\phi,f]$ contains a series of local actions which are the coefficients
of an expansion in powers of~$\hbar$:
\be
S[\phi,f] = S_0+ \hbar S_1 +\hbar S_2 +O(\hbar^3) \; \; , \; \; \; \;
S_i=\int dx {\cal L}_i \; \; ,
\ee
where $f$ stands for a set of external fields.
Accordingly, the generating functional $Z[f]$ can be calculated as an
expansion in $\hbar$:
\be
Z[f] = Z_0+ \hbar Z_1 +\hbar Z_2 +O(\hbar^3) \; \; ,
\ee
where $Z_0=S_0$. 
While for $Z_1$ we know an explicit, closed expression (in terms of the
logarithm of the determinant of a differential operator), that also
makes the calculation of the divergent part trivial, the calculation of
$Z_2$ is much more cumbersome.
It requires the calculation of the diagrams shown in Fig. \ref{fig:2loop},
whose exact meaning we cannot define here for lack of space. We refer the
interested reader to (\cite{JO}). A complete description of this calculation
and of various subtleties connected to the renormalization at order $p^6$
in CHPT (already discussed in part also in \cite{BCEGS}) will be given in a
forthcoming article (\cite{BCE}).

In such lengthy calculations, any kind of check is most welcome. In this
case the main one is: all divergences that are not polynomials in the
external fields must cancel in the end. These kind of divergences are
however present in single diagrams, and therefore the fact that they must
cancel when all diagrams are summed up represents a very thorough check on
the calculation.

The results of this calculation are best expressed in terms of a basis at
order $p^6$. Such a basis has been constructed by
\cite{FS} for the $SU(3)\otimes SU(3)$ case.  
Since we wanted to express our results for a generic number of light
flavours $N$ and also for $N=3$ and 2, we have worked out our own basis (in
fact two different versions of it). Our final list contains 115 independent
terms in the general $N$ case.
We are implementing all possible trace relations for the $N=3$ and $N=2$
case and  will then be able to compare to the basis chosen by \cite{FS}.

\section{Conclusions and Acknowledgments}
The discussions in this working group show that there is progress and
good future prospects both in theory and experiment in this area.
We thank the organizers for a well run and efficient
meeting. F.G. and G.I. thank the organizers for providing financial
support for the attendance of this workshop.
%
%


\begin{thebibliography}
%
%
\bibitem{}{CPLEAR}{Adler et al. (1997a)}
     Adler, R. et al., CPLEAR Collaboration (1997): 
     Phys. Lett. B407 (1997) 193. 
\bibitem{}{E78712}{Adler et al. (1997b)}
     Adler, S. et al., E787 Collaboration (1997b): Preprint 
     BNL-64627, hep-ex/9708012.
\bibitem{}{adler1997c}{Adler et al. (1997c)}
Adler,~S. et al., E787 Collab. (1997c), Phys. Rev. Lett. 79 (1997) 2204
\bibitem{}{Alemany}{Alemany (1997)}
Alemany,~R. M. Davier and A. H\"ocker, LAL 97-02, hep-ph/9703220.
\bibitem{}{ALEPH}{Aleph (1997)}
ALEPH Coll., CERN-PPE/97-013, January 28, 1997.
\bibitem{}{Roberts}{Alkofer and Roberts (1996)}
Alkofer,~R. and C.~Roberts, Phys. Lett. B369(1996) 101
\bibitem{}{Amsler}{Amsler (1997)}
Amsler,~C. hep-ex/9708025
\bibitem{}{anomaly}{Anomaly (1985)}
Anomaly (1985):
Donoghue,~J., Holstein,~B, Lin,~Y: Phys. Rev. Lett. 55 (1985) 2766\\
Bijnens,~J., Bramon,~A, Cornet,~F.: Phys. Rev. Lett. 61 (1988)\\
Bijnens,~J., Bramon,~A, Cornet,~F.: Z. Phys. C46 (1990) 599\\
Donoghue,~J., Wyler,~D.: Nucl. Phys. B316 (1989) 289\\
Ametller,~Ll., Bijnens,~J., Bramon,~A., Cornet,~F: Phys. Rev. D45 (1992) 986\\
Bijnens,~J.: Int. J. Mod. Phys. A8 (1993) 3045
\bibitem{}{antipov}{Antipov et al. (1987)}
Antipov,~Y.N. et al., Phys. Rev. D36 (1987) 21
\bibitem{}{Antonelli}{Antonelli et al. (1996)}
     Antonelli, V. et al.:
     Nucl. Phys. B469 (1996) 181.
\bibitem{}{CERN}{Bailey (1977)}
Bailey,~J. et al., Phys. Lett. B 68 (1977) 191;
F.J.M. Farley, E. Picasso, in ``Quantum Electrodynamics'', ed. T. Kinoshita,
World Scient. 1990.
\bibitem{}{PDG}{Barnett et al. (1996)}
     Barnett, R. et al. (1996): {\it Review of Particle Properties}, 
     Phys. Rev. D54 (1996) 1.
\bibitem{}{BMS83}{Bergstr\"om et al. (1983-1990)}
     Bergstr\"om, L.,  Mass\'o, E., Singer, P. (1983-1990):
     Phys. Lett. B131 (1983) 229; 
     Phys. Lett. B249  (1990) 141.
\bibitem{}{WG1}{Bernstein and Kaiser (1997)}
Bernstein,~A., Kaiser,~N. (1997): Working group on Electromagnetic Production
of Goldstone Bosons, these proceedings
\bibitem{}{BBC}{Bijnens, Bramon and Cornet (1990)}
Bijnens,~J. A.~Bramon and F.~Cornet, Phys. Lett. B237 (1990) 488
\bibitem{}{Bijnens}{Bijnens (1997)}
Bijnens,~J. (1997): these proceedings
\bibitem{}{BP}{Bijnens and Prades (1994)}
Bijnens,~J., Prades,~J.: Z. Phys. C64 (1994) 475
\bibitem{}{BPP}{Bijnens et al. (1995)}
Bijnens,~J. E. Pallante and J. Prades (1995), Phys. Rev. Lett. 75 (1995) 1447;
75 (1995) 3781 (E); Nucl. Phys. B474 (1996) 379.
\bibitem{}{BT}{Bijnens and Talavera (1997)}
Bijnens,~J., Talavera,~P.: Nucl. Phys. B489 (1997) 387
\bibitem{}{BCEGS}{Bijnens et al. (1997a)}
Bijnens,~J. G.~Colangelo, G.~Ecker, J.~Gasser and M.~Sainio (1997a),
  hep-ph/9707291, Nucl. Phys. B, in press.
\bibitem{}{BCE}{Bijnens et al. (1997b)}
J.~Bijnens, G.~Colangelo and G.~Ecker (1997b), work in progress.
\bibitem{}{BO+95}{Bondar et al. (1995)}
Bondar,~B. et al., Phys. Lett. B356 (1995) 8 
\bibitem{}{CLEOeta}{T. Browder et al. (1997)}
Browder,~T. et al, CLEO collab.,  CLNS-97-1484, June 1997, hep-ex/9706005
\bibitem{}{BuBu}{Buchalla and  Buras (1994)} 
     Buchalla, G., Buras, A.J. (1994): Nucl. Phys. B412 (1994) 106.
\bibitem{}{BuBu96}{Buchalla and Buras (1996)}     
     Buchalla, G., Buras, A.J. (1996): 
     Phys. Rev. D54 (1996) 6782. 
\bibitem{}{Buras}{Buras et al. (1994)}
     Buras, A.J. et al.
     (1994):   Nucl. Phys., B423 (1994) 349.
\bibitem{}{CA+96a}{Cal\'en et al. (1996)} 
Cal\'{e}n,~H. et al., Nucl. Instr. Meth.  A379 (1996) 57.  
\bibitem{}{CA+96b}{Cal\'en et al. (1997a)} 
Cal\'en,~H. et al., Phys.Rev.Lett. 79 (1997)2642.
\bibitem{}{CA+97}{Cal\'en et al. (1997b)}
Cal\'en,~H. et al., preprint TSL/ISV-97-0181.
\bibitem{}{DE87b}{Cappiello and D'Ambrosio (1988)}     
     Cappiello, L., D'Ambrosio, G. (1988):
     Nuovo Cimento 99A (1988) 155.
\bibitem{}{CD93}{Cappiello et al. (1993)}
     Cappiello, L., D'Ambrosio, G., Miragliuolo, M. (1993):
     Phys. Lett. B298 (1993) 423. 
\bibitem{}{CLEO}{CLEO}
CLEO Collaboration, CLNS 97/1477, hep-ex/9707031
\bibitem{}{weak}{Czarnecki}
Czarnecki,~A. B. Krause and W. Marciano, Phys. Rev. D52 (1995) R2619;
Phys. Rev. Lett. 76 (1996) 3267;\\
S. Peris, M. Perrottet and E. de Rafael, Phys. Lett. B355 (1995) 523;\\
 A. Czarnecki, B. Krause, hep-ph/9606393.\\
T.V. Kukhto et al., Nucl. Phys. B371 (1992) 567.
\bibitem{}{CE93}{Cohen et al. (1993)}
     Cohen, A.G., Ecker, G., Pich, A. (1993):  Phys. Lett. B304 (1993) 347.
\bibitem{}{DIPP}{D'Ambrosio et al. (1994)}
     D'Ambrosio, G. et al. (1994):
     Phys. Rev. D50 (1994) 5767; (E) D51 (1994) 3975. 
\bibitem{}{DEIN95}{D'Ambrosio et al. (1995)} 
     D'Ambrosio,~G. et al. (1995): 
in {\it The Second DA$\Phi$NE Physics Handbook}, Eds. Maiani, L.
     Pancheri, G., Paver, N. (Frascati, 1995), p. 265.
\bibitem{}{DEIN96}{D'Ambrosio et al. (1996)}
     D'Ambrosio, G. et al. (1996):
 Phys. Lett. B380 (1996) 165;
     Z. Phys. C76 (1997) 301.
\bibitem{}{DI96}{D'Ambrosio and Isidori (1996)} 
     D'Ambrosio, G., Isidori, G. (1996): 
 Preprint INFNNA-IV-96/29, hep-ph/9611284.
\bibitem{}{DP96} {D'Ambrosio and Portol\'es (1996)} 
     D'Ambrosio, G., Portol\'es, J. (1996):
     Phys. Lett. B389 (1996) 770; (E) B395 (1997) 389.
\bibitem{}{DPN}{D'Ambrosio and  Portol\'es (1997)}
     D'Ambrosio, G., Portol\'es, J. (1997): Nucl. Phys. B492 (1997) 417.
\bibitem{}{DIP}{D'Ambrosio et al. (1997)}
     D'Ambrosio, G.,  Isidori, G.,  Portol\'es, J. (1997):
   preprint  INFNNA-IV-97/40, hep-ph/9708326.
\bibitem{}{DMO}{Das, Mathur and Okubo (1967)}
Das,~T., V.S.~Mathur and S.~Okubo, Phys. Rev. Lett. 19 (1967) 859
\bibitem{}{DG95}{Donoghue and Gabbiani (1995)} 
     Donoghue, J.F., Gabbiani, F. (1995): Phys. Rev. D54 (1995) 2187.
\bibitem{}{DG97}{Donoghue and Gabbiani (1997)}
     Donoghue, J.F., Gabbiani, F. (1997): Phys. Rev. D56 (1997) 1605.
\bibitem{}{DG}{Donoghue and Golowich (1994)}
Donoghue,~J.F. and E.~Golowich, Phys. Rev. D49 (1994) 1513
\bibitem{}{WG3}{Drechsel (1997)}
Drechsel,~D.(1997): Working Group on Hadronic Form Factors, these proceedings
\bibitem{}{DE87a}{Ecker et al. (1987a)}
     Ecker, G., Pich, A., de Rafael, E. (1987a): 
     Phys. Lett. B189 (1987) 363.
\bibitem{}{EPR1}{Ecker et al. (1987b)}
     Ecker, G., Pich, A., de Rafael, E. (1987b):
     Nucl. Phys. B291 (1987) 692.
\bibitem{}{EPR88}{Ecker et al. (1988)} 
     Ecker, G., Pich, A., de Rafael, E. (1988): 
     Nucl. Phys. B303 (1988) 665.
\bibitem{}{EP90}{Ecker et al. (1990)}
     Ecker, G., Pich, A., de Rafael, E. (1990): 
     Phys. Lett. B237 (1990) 481.
\bibitem{}{EKW}{Ecker et al. (1993)}
     Ecker, G., Kambor, J., Wyler, D. (1993): Nucl. Phys. B394 (1993) 101.
\bibitem{}{Eckerrep}{Ecker (1995)}
     Ecker, G. (1995):
     Prog. Part. Nucl. Phys. 35 (1995) 1.
\bibitem{}{Ecker}{Ecker (1997)}
Ecker,~G. (1997): these proceedings.
\bibitem{}{FS}{Fearing and Scherer (1996)}
Fearing,~H. and S.~Scherer, Phys. Rev. D53 (1996) 315.
\bibitem{}{FNR}{Floratos, Narison and Pich (1979)} 
Floratos,~E.G, S.~Narison and E.~de~Rafael,
Nucl. Phys. B155 (1979) 115 
\bibitem{}{Gasser}{Gasser (1997)}
Gasser,~J. (1997): these proceedings
\bibitem{}{GL}{Gasser and Leutwyler (1985)}
Gasser,~J. and H.~Leutwyler, Nucl. Phys. B250 (1985) 465
\bibitem{}{georgi}{Georgi (1994)}
Georgi,~H.
Phys.Rev. D49 (1994) 1666
\bibitem{}{GK1}{Golowich and Kambor (1995)}
Golowich,~E. and J.~ Kambor, Nucl. Phys. B447 (1995) 373
\bibitem{}{GK2}{Golowich and Kambor (1996)}
Golowich,~E. and J.~Kambor, Phys. Rev. D53 (1996) 2651
%
\bibitem{}{GK3}{Golowich and Kambor (1997a)}
Golowich,~E. and J.~Kambor, 'Chiral Sum Rules to Second Order in 
Quark Mass', Phys. Rev. Lett. (to be published); hep-ph/9797341 
%
\bibitem{}{GK4}{Golowich and Kambor (1997b)}
Golowich,~E. and J.~Kambor, 'Two-loop Analysis of Axialvector Current
Propagators in Chiral Perturbation Theory', hep-ph/9710214
\bibitem{}{HKS}{Hayakawa (1995)}
Hayakawa,~M. T. Kinoshita and A. Sanda (1995), Phys. Rev. Lett. 75 (1995) 790;
Phys. Rev. D54 (1996) 3137;\\
M. Hayakawa and T. Kinoshita, KEK-TH530, hep-ph/9708227
\bibitem{}{BNL}{Heinson et al. (1995)}
     Heinson, A.P. et al., E791 Collaboration (1995):
 Phys. Rev. D51 (1995) 985.
\bibitem{}{Taronetal}{Herrera-Siklody et al. (1997)}
Herrera-Siklody,~P. et al.:
Nucl.Phys.B497 (1997) 345 and hep-ph/9710268 
\bibitem{}{Holstein}{Holstein (1996)}
Holstein,~B Phys. Rev. D53 (1996) 4099 
\bibitem{}{JO}{Jack and Osborn (1982)}
Jack,~I. and H.~Osborn, Nucl. Phys. B207 (1982) 474.
\bibitem{}{eedata}{Jegerlehner 1996}
Jegerlehner,~F. hep-ph/9606484, and refs. therein;\\
S. Eidelman, F. Jegerlehner, Z. Phys. C67 (1995) 585;\\
T. Kinoshita, B, Nizi\'c, Y. Okamoto, Phys. Rev. D 31 (1985) 2108;\\
W.A. Worstell, D.H. Brown, Phys. Rev. D 54 (1996) 3237.
\bibitem{}{OO}{Kaiser et al. (1995)}  Kaiser,~N. P. B. Siegel and W. Weise,
{\em Nucl. Phys.}{\bf A594}  325 (1995).
J. A. Oller and E. Oset, {\em Nucl. Phys.}{\bf A620} 438(1997).
\bibitem{}{KMW89}{Kambor et al. (1989)}
     Kambor, J., Missimer, J., Wyler, D. (1989):
 Nucl. Phys. B346 (1990) 17.
\bibitem{}{KMW91}{Kambor et al. (1991)}
     Kambor, J., Missimer, J., Wyler, D. (1991):
 Phys. Lett. B261 (1991) 496. 
\bibitem{}{KDHMW}{Kambor et al. (1992)}
     Kambor, J., Donoghue, J.F., Holstein, B.R.,  Missimer, J., 
     Wyler, D. (1992): Phys. Rev. Lett. 68 (1992) 1818.
\bibitem{}{KH94}{Kambor and Holstein (1994)}
     Kambor, J., Holstein, B.R. (1994):
     Phys. Rev.  D49 (1994) 2346.
\bibitem{}{KNO}{Kinoshita (1985)}
Kinoshita,~T. B. Nizi\'c, Y. Okamoto, Phys. Rev. D 31 (1985) 2108.
\bibitem{}{QED}{Kinoshita (1996)}
Kinoshita,~T. Rep. Prog. Phys. 59 (1996) 1459.
\bibitem{}{TAK96}{Kitching et al. (1997)} 
     Kitching, P. et  al.,  E787 Collaboration (1997):
     BNL-64628, hep-ex/9708011.
\bibitem{}{Krause}{Krause}
Krause,~B. hep-ph/9607259.
\bibitem{}{BNLm}{Lee Roberts (1992)}
Lee Roberts,~B. Z. Phys. C 56 (Proc. Suppl.) (1992) 101;\\ 
see also {\tt http://www.phy.bnl.gov/g2muon/home.html}
\bibitem{}{ASY}{Lepage (1979)}
Lepage,~G. S.J. Brodsky, Phys. Lett. B 87 (1979) 359;\\
Manohar,~A.  Phys. Lett. B  244 (1990) 101;\\
J.M. G\'erard, T. Lahna,  Phys. Lett. B 356 (1995) 381.
\bibitem{}{leutwyler97}{Leutwyler (1996)}
Leutwyler,~H. Phys. Lett. B374 (1996) 181 and hep-ph/9709408 
\bibitem{}{litten89}{Littenberg (1989)}
     Littenberg, L. (1989): Phys. Rev. D39 (1989) 3322.
\bibitem{}{Lowe}{Lowe (1997)}
Lowe,~J. for the BNL E865 experiment, 1997.
\bibitem{}{WG2}{Mei{\ss}ner and Sevior (1997)}
Mei{\ss}ner,~U.-G., Sevior,~M. (1997): Working Group on
$\pi\pi$ and $\pi N$ interactions , these proceedings
\bibitem{}{Miskimen}{Miskimen, Wang and Yegneswaran (1994)}
Miskimen,~R. K.~Wang and A.~Yegneswaran (Spokespersons),
{\em Study of the Axial Anomaly using the $\gamma\pi^+\to\pi^+\pi^0$
Reaction near Threshold, TJNAF-E94015}
\bibitem{}{Moinester}{Moinester (1997)}
Moinester,~M (1997): these proceedings
\bibitem{}{SUSY}{Moroi}
Moroi,~T. Phys. Rev. D 53 (1996) 6565;\\
M. Carena, G.F. Giudice, C.E.M. Wagner, hep-ph/9610233.
\bibitem{}{NR}{Narison and de Rafael (1980)} 
Narison,~S. and E.~de~Rafael,
Nucl. Phys. B169 (1980) 253 
\bibitem{}{NEB77}{Novikov et al. (1977)}
     Novikov, V.A. et al.
     (1977):  Phys. Rev. D16 (1977) 223.
\bibitem{}{klpipigg}{O'Dell (1997)}
O'Dell,~V., KTeV collab., {\em Status and New Results of the KTeV Rare
Kaon Program}, MIST workshop, May 1997, Fermilab.
\bibitem{}{fut}{Oller et al. (1997)}
Oller,~J.A., E. Oset and J. R. Pel\'aez,
in preparation, 1997.
\bibitem{}{Pallante}{Pallante}
Pallante,~E. Phys. Lett. B341 (1994) 221
\bibitem{}{peris}{Peris (1994)}
Peris,~S. Phys. Lett. B324 (1994) 442
\bibitem{}{derafaelperis}{Peris and de Rafael (1995)}
Peris,~S. and E. de Rafael,
Phys. Lett. B348 (1995) 539
\bibitem{}{PI91}{Pich and de Rafael (1991)}
     Pich, A., de Rafael, E. (1991): Nucl. Phys. B358 (1991) 311.
\bibitem{}{deRafael}{de Rafael (1994)}
     de Rafael, E. (1994): {\it Chiral Lagrangians and kaon CP
     violation}, in {\it CP Violation and the limits of the 
     Standard Model} TASI94 Proceedings (Boulder, 1994).
\bibitem{}{Rafael}{de Rafael}
de Rafael,~E. Phys. Lett. B322 (1994) 239
\bibitem{}{Rosselet}{Rosselet et al. (1977)}
Rosselet,~L. et al., Phys. Rev. D15 (1977) 574
\bibitem{}{Stern}{Stern (1997)}
Stern,~J. (1997): these proceedings
\bibitem{}{IAM}{Truong (1988)} Truong,~T.N.
{\em Phys. Rev. Lett.} {\bf 61} (1988) 2526, ibid {\bf 67}, 2260 (1991);
A. Dobado, M. J. Herrero and T. N. Truong,
{\em Phys. Lett.} {\bf B235} 129 (1990); 
A. Dobado and J. R. Pel\'aez, {\em Phys. Rev.}{\bf D47},
4883 (1992), ibid {\em Phys. Rev.}{\bf D56} (1997) 3057;
T. Hannah, {\em Phys. Rev.}{\bf D52} 4971(1995), 
ibid {\bf D54} 4648 (1996), ibid{\bf D55} 5613 (1997) 
and this proceedings in \cite{WG3}.
\bibitem{}{Ueda}{Ueda (1991)}
Ueda,~T. Phys. Rev. Lett. 66 (1991) 297 and
Ueda,~T. Phys. Lett. B292 (1992) 228
\bibitem{}{Kolck}{van Kolck (1997)}
van Kolck~U. (1997):  these proceedings
\bibitem{}{VEPP2M}{VEPP 2M}
VEPP 2M, new unpublished data, Novosibirsk
\bibitem{}{Zou}{Zou et al. (1996)}
     Zou, Y. et al. (1996): Phys. Lett. B369 (1996) 362.
\end{thebibliography}
\end{document}